\documentclass{svproc}
\usepackage{url}
\usepackage{graphicx}
\usepackage{amsmath}
\usepackage{amssymb}

\begin{document}
\mainmatter              
\title{Mixed Membership Models for Multilevel Functional Data}
\titlerunning{Multilevel Functional Mixed Membership}  
%
\author{Donatello Telesca\inst{1} \and Nicholas Marco\inst{2} \and
Emma Landry\inst{1}}
\authorrunning{Donatello Telesca et al.} 
%
\tocauthor{Donatello Telesca, Nicholas Marco, and Emma Landry}
\institute{University of California Los Angeles, Los Angeles CA, USA,\\
\email{dtelesca@ucla.edu},
\and
Duke University,
Duram NC, USA}

\maketitle              

\begin{abstract}
Mixed membership models extend classical clustering by substituting the notion of uncertain membership with the notion of mixed membership. In particular, these models allow each observation to partially belong to multiple pure membership classes. We discuss mixed membership models for functional data by extending the framework to multilevel functional observations. We show how the classical multivariate Karhunen-Loève decomposition can be translated into a simple hierarchical model for scalable and flexible expressivity of the underlying stochastic processes. The identifiability of partial membership structures is aided by the definition of a hierarchical repulsive prior on the unitary simplex. Our work is motivated and illustrated by applications to a study on functional brain imaging through electroencephalography (EEG) of children with autism spectrum disorder (ASD). 
\end{abstract}
\section{Introduction}
\label{sec:intro}
Our work is motivated by electroencephalograpy (EEG) imaging studies of children with autism spectrum disorder (ASD).  Figure 1 illustrates log spectral densities from a resting state electroencephalogram (EEG), for a random sample of children diagnosed with ASD (ages 2–12) and age‑matched typically developing (TD) controls \cite{Dick1}. EEG data were obtained during a two‑minute recording session  on a 128‑channel HydroCel Geodesic Sensor Net interpolated to a 25‑channel montage.  Within subject, EEG spectral features differ slightly between channels, with differences observed in both shape and relative amplitude of the EEG spectrum. Much larger patterns of heterogeneity are observed between subjects, as each unit contributes EEG spectral data with different spectral features.      

\begin{figure}
\label{data}
\begin{center}
\begin{tabular}{cc}
\includegraphics[width=.49\textwidth]{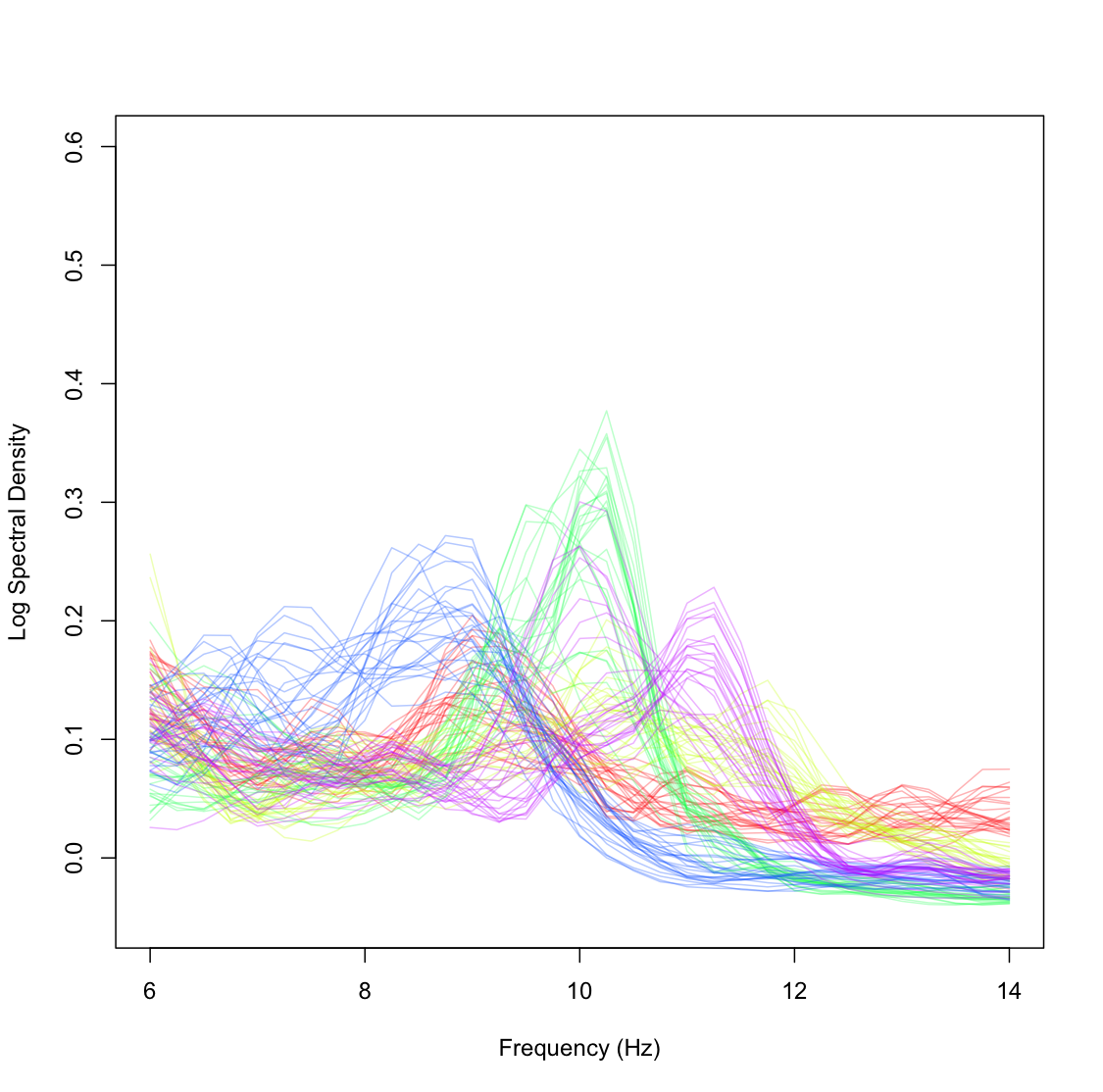} & \includegraphics[width=0.49\textwidth]{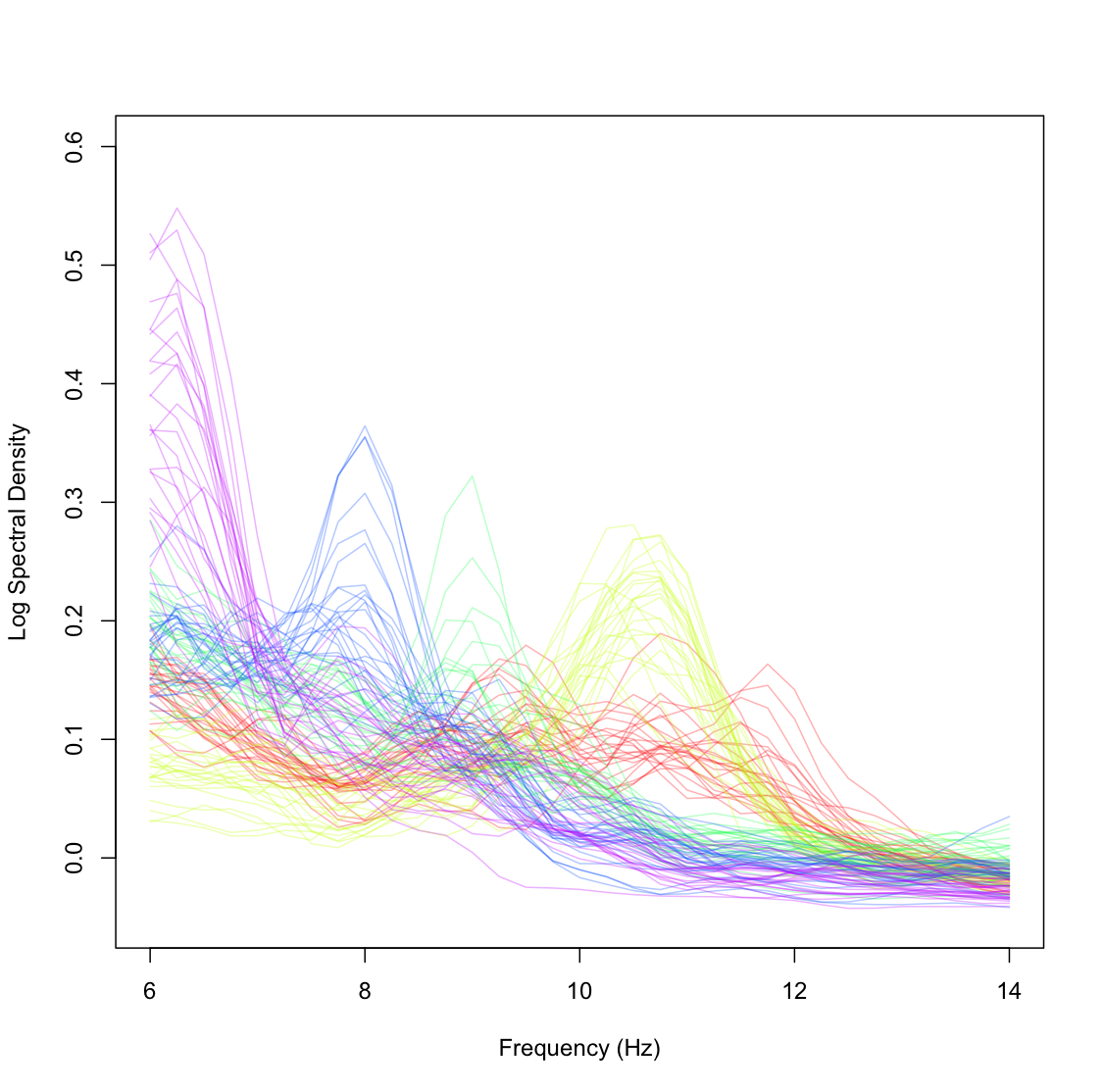}\\
\end{tabular}
\end{center}
\caption{Log spectral densities for a sample of five TD (right panel) and five ASD children (left panel) over 25 electrodes. Electrodes, within subject are coded to be expressed in the same color.}
\end{figure}

We conceptualize these data structures as multilevel functional data, i.e. data where each statistical unit contributes, say $J$ functions, s.t., given an evaluation domain $\mathcal{T}$, each datum $Y_i(t)$ belongs to a product space of measurable functions, i.e. $L(\mathcal{T}):= \mathcal{L}^{(1)}_2(\mathcal{T})\times\cdots\times \mathcal{L}^{(J)}_2(\mathcal{T})$, for $i=1,2,\ldots,n$. We note that this characterization is equivalent to the notion of multivariate functional data \cite{Happ}. 

In an unsupervised learning setting, we are interested in the case where observations possibly belong to multiple pure membership classes simultaneously, yielding mixed‑membership (or partial membership) models \cite{Erosheva}. 

We propose adapting the idea of mixed‑membership functions \cite{Marco24} to the multilevel setting, by assuming a known number of latent functional features, and defining a functional mixed‑membership process through mixtures of dependent Gaussian Processes (GP). To ensure flexible expressivity, we leverage the multivariate Karhunen–Loève construction of \cite{Happ} and assume mixed membership is organized hierachically within subject. Mixed membership identifiablity is facilitated by a hierarchically repulsive prior on the unit simplex. 


The remainder of this article is organized as follows. Section~\ref{Model} introduces the proposed multilevel mixed membership representation. Section~\ref{Prior} introduces priors and a discussion of posterior inference. An application to ASD data is introduced in Section~\ref{data} and we conclude with a critical discussion in Section~\ref{discussion}.

\section{Multilevel Mixed Membership of Functions}
\label{Model}

Let $\{\mathbf{Y}_i(\cdot)\}_{i=1}^N$ denote the observed sample paths that we represent as realization of a stochastic process defined in the product space defined earlier, $L(\mathcal{T})$. Since the observed sample paths are observed at only a finite number of points, we will let $\mathbf{t}_i = [t_{i1}, \dots, t_{in_i}]'$ denote the time points at which the $i^{th}$ function was observed. Accordingly, we observe only finite dimensional path evaluation $\mathbf{Y}_i(\mathbf{t}_i)\in \mathbb{R}^{n_i\times J}$, with univariate component functions $Y_{ij}(\mathbf{t}_i)\in \mathbb{R}^{n_i}$, ($j=1,2,\ldots, J$).  

We aim to define mixed membership of each multilevel functional path $Y_{ij}(\cdot)$ wrt. $K$ latent pure membership functional features. To do so, without loss of generality, we define the sampling model for the finite-dimensional margins of $Y_{ij}({\bf t}_i)$ as a convex combination of dependent GPs. Let $\mathbf{B}^\prime(t) := [b_1(t), b_2(t), \dots, b_P(t)]$ be a set of $P$ fixed basis functions for $\mathcal{T}$, i.e. B-splines, and define  $\mathbf{S}(\mathbf{t}_i) = [\mathbf{B}(t_1) \cdots \mathbf{B}(t_{n_i})] \in \mathbb{R}^{P \times n_i}$, we assume: 
\begin{equation}
    \label{eq: likelihood}
    Y_{ij}({\bf t}_i)\mid \boldsymbol{\Theta}, \mathbf{X} \sim \mathcal{N}\left\{ \sum_{k=1}^K Z_{ijk}\left(\mathbf{S}({\bf t}_i)^T \boldsymbol{\nu}_k  + \sum_{m=1}^M\chi_{ijm}\mathbf{S}({\bf t}_i)^T \boldsymbol{\phi}_{km}\right),\; \sigma_j^2 \mathbf{I}_{n_i}\right\},
\end{equation}
where  $\boldsymbol{\Theta}$ denotes the collection of the model parameters. The convex combination over latent functional features depends on subject by channel  mixing proportions $Z_{ijk}$, which are variables that lie on the unit simplex, such that $Z_{ijk} \in (0,1)$ and $\sum_{k=1}^K Z_{ijk} = 1$. 

Integrating out the pseudo-scores $\chi_{ijm}$, for $i = 1, \dots, N$, $j=1,\ldots, J$ and $m = 1, \dots, M$, we obtain the following finite dimensional margins:
\begin{equation}
    \label{eq: likelihoodIntChi}
    Y_{ij}(\mathbf{t}_i)\mid  \boldsymbol{\Theta}_{-\chi} \sim \mathcal{N}\left\{\sum_{k=1}^KZ_{ijk}\mathbf{S}(t_i)^T \boldsymbol{\nu}_k ,\; \mathbf{V}(\mathbf{t}_i, \mathbf{z}_i) + \sigma_j^2\mathbf{I}_{n_i}\right\},
\end{equation}
where $\boldsymbol{\Theta}_{-\chi}$ is the collection of the model parameters excluding the $\chi_{ijm}$ parameters ($i = 1, \dots, N$ $j=1,2,\ldots, J$, and $m = 1, \dots, M$) and $$\mathbf{V}(\mathbf{t}_i, \mathbf{z}_i) =  \sum_{k=1}^K\sum_{k'=1}^K Z_{ijk}Z_{ijk'}\left\{\mathbf{S}(\mathbf{t}_i)^T\sum_{m=1}^{M}\left(\boldsymbol{\phi}_{km}\boldsymbol{\phi}_{k'm}\right)\mathbf{S}(\mathbf{t}_i)\right\}.$$
Equation \ref{eq: likelihoodIntChi} illustrates that the proposed  functional mixed membership model can be expressed as an additive model; the mean structure is a subject-by-channel convex combination of the feature-specific means, while the covariance can be written as a weighted sum of covariance functions and cross-covariance functions. 

To have an adequately expressive and scalable model, we approximate the covariance surface of the $K$ features using $M$ scaled \textit{pseudo-eigenfunctions}. In this framework, orthonormality will not be imposed on the $\mathbf{S}({\bf t}_i)^T \boldsymbol{\phi}_{km}$ parameters. From a modeling perspective, this allows us to operate outside of the confines of a Stiefel manifold, facilitating better Markov chain mixing and easier sampling schemes. Although direct inference on the eigenfunctions is no longer available, a formal analysis can still be conducted by reconstructing the posterior samples of the covariance surface and calculating eigenfunctions from the posterior samples.

\section{Priors, Repulsion and Posterior Inference}
\label{Prior}

 Adaptive regularization for the covariance functions  is achieved by using the multiplicative gamma process shrinkage prior proposed by \cite{Bhattacharya} to achieve adaptive regularized estimation of the covariance structure. Letting $\phi_{kpm}$ be the $p^{th}$ element of $\boldsymbol{\phi}_{km}$, we have the following:
$$\phi_{kpm}\mid \gamma_{kpm}, \tilde{\tau}_{mk} \sim \mathcal{N}\left(0, \gamma_{kpm}^{-1}\tilde{\tau}_{mk}^{-1}\right), \;\;\; \gamma_{kpm} \sim \Gamma\left(\nu_\gamma /2 , \nu_\gamma /2\right), \;\;\; \tilde{\tau}_{mk} = \prod_{n=1}^m \delta_{nk},$$
$$ \delta_{1k}\mid a_{1k} \sim \Gamma(a_{1k}, 1), \;\;\; \delta_{jk}\mid a_{2k} \sim \Gamma(a_{2k}, 1), \;\;\; a_{1k} \sim \Gamma(\alpha_1, \beta_1), \;\;\; a_{2k} \sim \Gamma(\alpha_2, \beta_2),$$
where $1 \le k \le K$, $1 \le p \le P$, $1 \le m \le M$, and $2 \le j \le M$. Assuming $\alpha_2 > \beta_2$, we obtain $\mathbb{E}(\tilde{\tau}_{mk}) > \mathbb{E}(\tilde{\tau}_{m'k})$ for $1 \le m < m' \le M$, leading to the prior on $\phi_{kpm}$ having stochastically decreasing variance as $m$ increases.

Smoothness of the feature means and pseudo-eigenfunctions, rely on the use a first-order random walk penalties on $\boldsymbol{\nu}_k$ and $\boldsymbol{\phi}_{mk}$. 
We also assume conditionally conjugate Inverse Gamma priors for the error variances, s.t. $\sigma_j^2 \sim_{iid} IG(\alpha_0,  \beta_0)$.

Mixed membership models face multiple identifiability problems due to their increased flexibility over traditional clustering models \cite{Marco24} \cite{Chen23}. 
Therefore, geometric constraints must be placed on the allocation parameters $\mathbf{Z}$ to ensure identifiabilty of the the mixed membership structure. 
One of the weakest geometric conditions is known as the \textit{sufficiently scattered} condition \cite{Marco24} \cite{Chen23}. This constraint  requires the collection of membership probabilities $\mathbf{Z}$ to cover the $K$-unit simplex sufficiently well - convex hull closely reaching every corner and edge.

We propose a stochastic constraint on the distribution of $\mathbf{Z}$ by constructing a \emph{hierarchically repulsive point-process prior}. Specifically, let $\{\boldsymbol{\pi}_i\}_{i=1}^n$, be $n$-realizations of a point process in the $K$-unit simplex.  Following \cite{Beraha22}, a repulsive process may be defined as follows,
\begin{equation}
   P\left(\boldsymbol{\pi}_{1:n} \mid \boldsymbol{\alpha}, \tau\right) 
= \frac{1}{C_\pi} 
\underbrace{\prod_{i=1}^n \Bigg( \frac{1}{B(\boldsymbol \alpha)} 
   \prod_{k=1}^K \pi_{ik}^{\alpha_{k} -1}\Bigg)}_{\text{Dirichlet distribution}}
\;
\underbrace{\prod_{1 \leq i ,j \leq N} \Bigg[ \exp \left( - \frac{\tau}{\rVert \boldsymbol \pi_i - \boldsymbol \pi_j \rVert^2}\right)\Bigg]^{1/N}}_{\text{repulsion}},
\end{equation}
where, $C_\pi$ is a normalizing constant, and $\tau$ is a hyperparameter controlling the repulsiveness.  Given subject-level repulsive centroids $\boldsymbol{\pi}_i$, subject-by channel mixing probabilities are assumed 
$${\bf Z}_{ij}\mid \boldsymbol{\pi}_i, \eta \sim_{iid} \mbox{Dirichlet}(\eta\,\boldsymbol{\pi}),$$
with $\eta$ controlling channel-specific mixture variance around subject-level repulsive centroids $\boldsymbol{\pi}$.

Given the sampling and the prior models, the posterior measure is accessible via standard MCMC strategies and posterior simulation follows closely \cite{Marco24}.

\begin{figure}[h]
\label{fig:results}
\begin{center}
\begin{tabular}{cc}
\includegraphics[width=.4\textwidth]{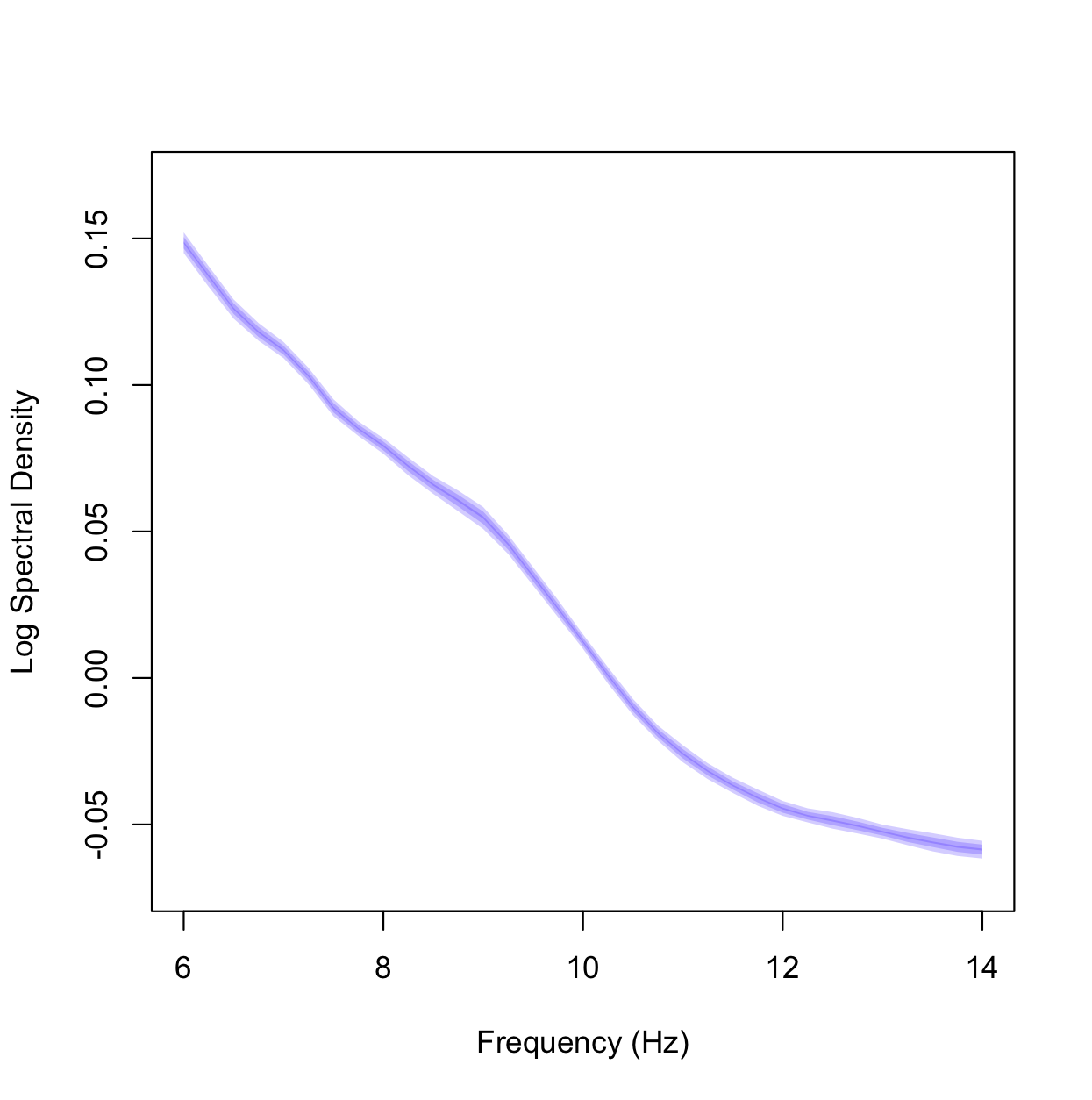} & \includegraphics[width=0.4\textwidth]{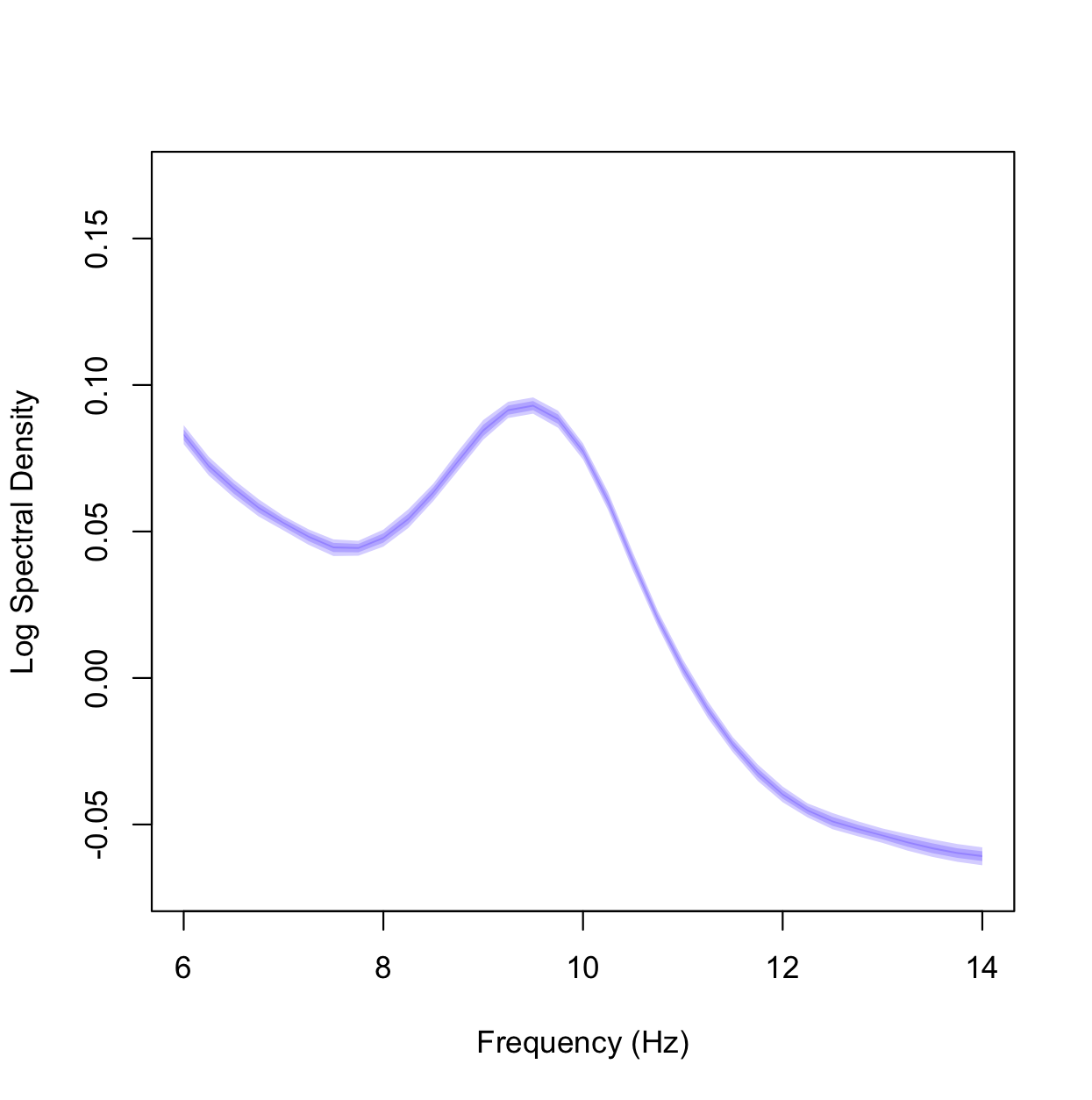}\\
\multicolumn{2}{c}{\includegraphics[width=0.8\textwidth]{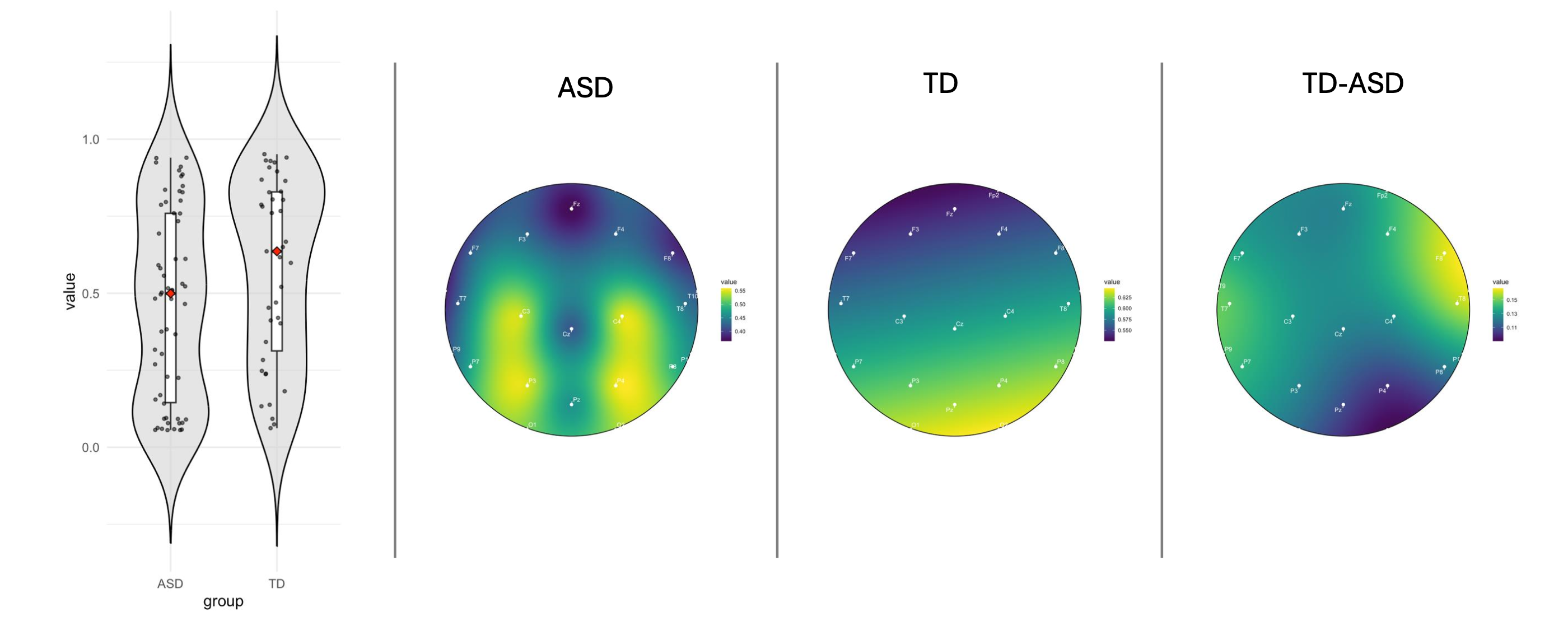}}
\end{tabular}
\end{center}
\caption{Mixed membership log spectral features (top panels). Population mixing proportions loading onto the alpha peak feature (bottom -left panel). Mean alpha peak loading intensity by EEG channel and TD to ASD difference (bottom-right panels).}
\end{figure}

\section{Peak Alpha Frequency and Autism Spectrum Disorder}
\label{data}
We consider an imaging study in ASD. The study includes EEG data (58 ASD, 39 typically developing children, ages 2–12) previously analyzed by \cite{Marco24}. We consider all 25 EEG channels in the alpha band (6–14 Hz). Clinically important is the peak alpha frequency (PAF): in typically developing children the alpha peak becomes more prominent and shifts to higher frequencies with age, whereas this pattern is attenuated in most children with ASD. 
In this context, functional clustering typically produces poorly separated and hard-to-interpret cluster means. 

We fit a mixed‑membership model with two functional features (K = 2; selected by AIC/BIC). Results are shown in Figure 2. This parsimonious choice yields: feature 1, an aperiodic 1/f (pink) noise–like mean with no alpha peak, and feature 2, a periodic component with a clear alpha peak typical of older neurotypical subjects. These two features separate periodic (alpha) and aperiodic (1/f) spectral patterns that coexist in EEG. Because subjects exhibit continuous mixtures of these patterns, mixed membership is more appropriate than hard clustering for describing the observed heterogeneity.

The subject level posterior mean peak-alpha feature loadings shows that TD children are highly likely to heavily load on feature 2 (well-defined PAF), whereas ASD children exhibit a higher level of heterogeneity. Accessing channel-level mixing probabilities, we obtain a clear view of which brain region exhibits a well defined alpha peak, with TD to ASD differentials highlighting meaningful difference in temporal regions of the brain. 

In general, our findings confirm related evidence in the
scientific literature on developmental neuroimaging but offer
a completely novel point of view in quantifying group mem-
bership as part of a spectrum.

\section{Discussion}
\label{discussion}
This paper extends the class of functional mixed‑membership models to multilevel functional data. Crucially, multilevel inference is conducted at the level of the mixing proportions, via a hierarchically repulsive point process on the unit-simplex.

Our work relies on defining a mixed‑membership Gaussian process via projections onto a basis subspace, using the multivariate Karhunen–Loève formulation \cite{Happ}. The construction yields a coherent sampling model and easily accessible posterior measure via standard MCMC simulation.

Model selection via information criteria tended to overestimate K in our experiments; the elbow method performed better. Nonparametric alternatives (e.g., Indian Buffet Process) exist but their operating characteristics and inference across changing dimensions are not well understood. Existing results on posterior behavior for overfitted mixture models, however,  do not directly extend to our case. Consequently, further work is needed on posterior convergence for overfitted mixed‑membership models.

%
%


\begin{thebibliography}{6}
%
\bibitem{Dick1}
Dickinson, A., DiStefano, C., Senturk, D., Jeste, S. S.: Peak alpha frequency is a neural marker of cognitive function across the autism spectrum. European Journal of Neuroscience, Vol. 47(6), pp. 643-651 (2018).

\bibitem{Li20}
Li, Q., Shamshoian, J., S¸ent¨urk, D., Sugar, C., Jeste, S., DiStefano, C., Telesca, D., et al.: Region-referenced spectral power dynamics of EEG signals: A hierarchical modeling approach. Annals of Applied Statistics 14, 2053–2068 (2020).

\bibitem{Happ}
Happ, C. and Greven, S: Multivariate functional principal component analysis for data observed on different (dimensional) domains. Journal of the American Statistical Association 113, 649–659 (2018).

\bibitem{Marco24}
Marco, N., Şentürk, D., Jeste, S., DiStefano, C., Dickinson, A., and Telesca, D.: Functional Mixed Membership Models. Journal of Computational and Graphical
Statistics, 33(4), 1139-1149. (2024).

\bibitem{Erosheva}
Erosheva, E., Fienberg, S., and Lafferty, J.: Mixed Membership
Models of Scientific Publications, PNAS. 101, 5220–5227 (2004).

\bibitem{Bhattacharya}
Bhattacharya, A., and Dunson, D. B.: Sparse Bayesian Infinite Factor Models. Biometrika, 98, 291–306 (2011).

\bibitem{Chen23}
Chen, Y., He, S., Yang, Y., and Liang, F: Learning Topic Models: Identifiability and Finite-Sample Analysis. Journal of the American Statistical Association, 118, 2860–2875 (2023).

\bibitem{Beraha22}
Beraha, M., Argiento, R., Møller, J., and Guglielmi, A.: MCMC Computations for Bayesian Mixture Models Using Repulsive Point Processes. Journal of Computational and Graphical Statistics, 31, 422–435 (2022).

\end{thebibliography}
\end{document}